\documentclass[useAMS,usenatbib,usegraphicx]{mn2e}
\usepackage{times}

\newcommand{\betaGLX}{\mbox{${\; {\rm \beta_{GLX}}}$}}
\newcommand{\betaOAO}{\mbox{${\; {\rm \beta_{OAO}}}$}}
\newcommand{\betaANS}{\mbox{${\; {\rm \beta_{ANS}}}$}}
\newcommand{\betaUIT}{\mbox{${\; {\rm \beta_{UIT}}}$}}
\newcommand{\betaCKS}{\mbox{${\; {\rm \beta_{CKS}}}$}}
\newcommand{\betaCKSshort}{\mbox{${\; {\rm \beta_{CKS}'}}$}}

\newcommand{\IRXUV}{\hbox{IRX-UV}}
\newcommand{\dpp}{\hbox{$d_{\rm P}$}}

\newcommand{\Ffuv}{\hbox{$F_{\rm FUV}$}}

\newcommand{\ldust}{\hbox{$L_{\rm dust}$}}

\newcommand{\ffuv}{\hbox{$\overline{f}_{{\rm FUV}}$}}
\newcommand{\fnuv}{\hbox{$\overline{f}_{{\rm NUV}}$}}
\newcommand{\hii}{\hbox{H{\sc ii}}}

\newcommand{\ldustfuv}{\hbox{$L_{\rm dust}/L_{\rm FUV}$}}

\newcommand{\lhahb}{\hbox{$L_{{\rm H}\alpha}/L_{{\rm H}\beta}$}}

\def\f60100{$f_{\nu}(60\,\mu{\rm m})/f_\nu(100\,\mu{\rm m})$}

\newcommand{\hda}{\hbox{H$\delta_{\rm A}$}}
\newcommand{\dn}{\hbox{D$_n(4000)$}}
\newcommand{\ha}{\hbox{H$\alpha$}}
\newcommand{\ewha}{\hbox{EW(H$\alpha$)}}
\newcommand{\hb}{\hbox{H$\beta$}}

\newcommand{\hahb}{\hbox{H$\alpha$/H$\beta$}}

\newcommand{\tp}{\hbox{$t_0$}}
\newcommand{\mstar}{\hbox{$M_\ast$}}

\newcommand{\Afuv}{\hbox{$A_{\rm FUV}$}}
\newcommand{\tauV}{\hbox{$\hat{\tau}_V$}}
\newcommand{\srel}{\hbox{$s_{\rm rel}$}}

\title[Star formation history and dust attenuation in 
galaxies drawn from ultraviolet surveys]
{Star formation history and dust attenuation in
galaxies drawn from ultraviolet surveys}

\author[X. Kong et al.]
{X. Kong$^{1,2}$,
S. Charlot$^{1,3}$\thanks{E-mail: charlot@iap.fr},
J. Brinchmann$^1$ and
S. M. Fall$^{4}$
\\
$^{1}$Max-Planck Institut f\"ur Astrophysik, 
Karl-Schwarzschild-Strasse
1, D-85748 Garching, Germany\\
$^{2}$Center for Astrophysics, University of Science and Technology
of China, 230026 Hefei, PR China\\
$^{3}$Institut d'Astrophysique de Paris, CNRS, 98 bis Boulevard
Arago, 75014 Paris, France\\
$^{4}$Space Telescope Science Institute, 3700 San Martin Drive, 
Baltimore, Maryland 21218, USA
}

\begin{document}

\date{MNRAS, in press}

\pagerange{\pageref{firstpage}--\pageref{lastpage}} \pubyear{2003}

\maketitle

\label{firstpage}

\begin{abstract}
We compile a new sample of 115 nearby, non-Seyfert
galaxies spanning a wide range of star formation activities, from 
starburst to nearly dormant, based on ultraviolet observations with
various satellites. We combine these observations with
infrared observations to study the relation between ratio of total 
far-infrared to ultraviolet luminosities and ultraviolet spectral slope
(the `\IRXUV' relation). We show that, at fixed ultraviolet spectral
slope, quiescent star-forming galaxies in our sample have systematically
lower ratio of total far-infrared to ultraviolet luminosities than
starburst galaxies. The strengths of spectral indices sensitive to
star formation history, such as the 4000\,{\AA} spectral discontinuity 
and the \ha\ emission equivalent width, correlate well with distance
from the mean relation for starburst galaxies in the \IRXUV\ diagram, 
while there is little or no correlation between the dust-sensitive 
\hahb\ ratio and this distance. This is strong observational evidence
that the star formation history is relevant to the `second parameter'
affecting the \IRXUV\ relation. We show that these results can be 
understood in the framework of the simple model of 
\citealt{2000ApJ...539..718C} for the transfer of starlight through the 
interstellar medium in galaxies.  We confirm that, for starburst 
galaxies, the tight \IRXUV\ relation can be understood most simply as
a sequence in overall dust content. In addition, we find that the 
broadening of the relation for quiescent star-forming galaxies can be
understood most simply as a sequence in the ratio of present to 
past-averaged star formation rates. We use a library of Monte Carlo 
realizations of galaxies with different star formation histories and 
dust contents to quantify the accuracy to which the ultraviolet 
attenuation \Afuv\ of a galaxy can be estimated from either the ratio
of far-infrared to ultraviolet luminosities or the ultraviolet spectral
slope. We provide simple formulae for estimating \Afuv\ as a function
of either of these observational quantities and show that the accuracy
of these estimates can be significantly improved if some constraints
are available on the ratio of present to past-averaged star formation
rates.

\end{abstract}


\section{Introduction}

Much of what we know about the universe at high redshift arises
from observations of the rest-frame ultraviolet emission of distant
galaxies that is accessible from the ground at optical and 
near-infrared wavelengths. For this reason, significant effort
has been devoted over the past several years to improve our
understanding of the star formation history and dust attenuation
in galaxies drawn from ultraviolet surveys. Recent progress in this 
area has come mainly from studies of a sample of 57 nearby starburst 
galaxies, for which a wide range of
observations were compiled by \citet*{1999ApJ...521...64M}. 
In particular, for these galaxies there is a remarkably tight 
correlation between ratio of total far-infrared to ultraviolet 
luminosities, \ldustfuv, and ultraviolet spectral slope, $\beta$,
which is referred to as the `infrared excess (IRX)-ultraviolet 
(UV)' relation (here $\beta$ is defined by a power-law fit of the form
$f_\lambda\propto \lambda^\beta$ to the spectrum at wavelengths 
$1300\la\lambda\la 2600\,${\AA}; \citealt{1994ApJ...429..582C}).
\citet{1999ApJ...521...64M} find that the \IRXUV\ relation 
allows reliable estimates of the attenuation by dust at ultraviolet 
wavelengths, \Afuv, from either \ldustfuv\ or $\beta$. This relation,
if universal, represents our best hope to estimate the attenuation by
dust and hence the star formation rates of high-redshift galaxies, for
which the rest-frame optical and infrared emission cannot be observed 
at present.

The main uncertainty affecting the \IRXUV\ relation is that it has been
established only for starburst galaxies. Recently, 
\citet{2002ApJ...577..150B} has compiled a small sample of nearby, 
more quiescent galaxies drawn from ultraviolet observations with
various satellites (here we use the term `quiescent' to describe
the broad class of star-forming galaxies between active starburst and
dormant early-type galaxies). His results suggest that quiescent 
galaxies deviate from the \IRXUV\ relation for starburst galaxies, in
the sense that quiescent galaxies tend to have `redder' ultraviolet
spectra at fixed \ldustfuv. This implies that the
simple \citet{1999ApJ...521...64M} recipe to estimate \Afuv\
from either \ldustfuv\ or $\beta$ may not hold for all galaxy types.
The results of \citet{2002ApJ...577..150B}
raise several questions. Can star formation 
history alone account for the distribution of galaxies in the \IRXUV\
diagram? Or do the galaxies have different dust properties?  
\citet{2000ApJ...539..718C}
have developed a simple dust model, based on an idealized 
prescription of the main features of the (clumpy) interstellar 
medium (ISM), which allows one to interpret the \IRXUV\ relation 
and the emission-line properties of starburst galaxies in terms of 
star formation rate and dust content. This model accounts for the 
different attenuation affecting young and old stars in a galaxy,
because of the dispersal of the (giant molecular) clouds in which 
stars form. Using this model, \citet{2000ApJ...539..718C} show
that the tight \IRXUV\ relation for starburst galaxies can be 
understood most simply as a sequence in the overall dust content.
Clearly, to gain insight into the origin of the \IRXUV\ relation for
different types of galaxies, we require further observations and 
modelling of not only the ultraviolet and infrared luminosities,
but also other observable quantities with distinct dependences on
dust content and star formation history.

In this paper, we compile a new sample of 115 nearby, non-Seyfert
galaxies spanning a wide range of star formation activities, from
starburst to nearly dormant, to explore the physical parameters driving
the location of a galaxy in the \IRXUV\ diagram. Our sample is drawn
from ultraviolet observations with various satellites. We combine these
observations with infrared observations and, for a subset of the 
galaxies, with optical spectroscopy from various sources. We examine, in
particular, the 4000~\AA\ spectral discontinuity and the \ha\ emission
equivalent width, which are good indicators of star formation activity,
and the \hahb\ ratio, which is a good indicator of attenuation by dust.
Based on these data alone, it appears that the broadening of the \IRXUV\
relation for quiescent star-forming galaxies is most likely driven by
star formation history. We combine the simple dust model of 
\citet{2000ApJ...539..718C} with the population synthesis code of 
\citet{2003MNRAS.344.1000B} and show that this model can account for the
ultraviolet and far-infrared properties of the galaxies in our sample.
The model allows us to understand why the ratio of present to
past-averaged star formation rates is the most likely `second' parameter
that, in combination with dust content, determines the position of a
galaxy in the \IRXUV\ diagram. In Section 2 below, we compile our galaxy
sample and investigate the origin of the \IRXUV\ relation from an
observational viewpoint. In Section 3, we interpret the \IRXUV\ relation
using the model. We explore the implications of our results for
estimates of the ultraviolet attenuation in galaxies in Section\,4. Our
conclusions are summarized in Section\,5.
	     
\section{The origin of the \IRXUV\ relation: observational insight}

In this section, we re-examine the \IRXUV\ relation from an
observational viewpoint. We combine ultraviolet observations from 
various satellites with observations from the {\it Infrared 
Astronomical Satellite} ({\it IRAS}) to compile a new
sample of nearby, non-Seyfert galaxies spanning a wide range of star
formation activities in the \IRXUV\ diagram.  We explore how the 
position of a galaxy in this diagram depends on observable properties
with distinct dependences on dust content and star formation history.

\subsection{The data}

We compile a new sample of nearby, non-Seyfert
galaxies spanning a wide range of star formation activities, using 
available observations with the {\it International Ultraviolet 
Explorer} ({\it IUE}), the {\it Orbiting Astronomical Observatory} 
({\it OAO}) and the {\it Astronomical Netherlands Satellite} 
({\it ANS}). We first extract all (non-Seyfert) spiral, elliptical,
irregular and emission-line galaxies (classes 80, 81, 82 and 88) from
the {\it IUE} Newly-Extracted Spectra (INES version 3.0) archive. The
spectra in this database extend over two wavelength domains,
corresponding to the far- (1150--1980~{\AA}) and near-ultraviolet
(1850--3350~{\AA}) spectrographs on-board {\it IUE}. We reject galaxies
with only near-ultraviolet spectra, for which the ultraviolet spectral
slope cannot be measured reliably. When both the far- and 
near-ultraviolet spectra are available for a galaxy, we merge them to
form a single ultraviolet spectrum encompassing the range from 1150
to 3350~\AA\ (the spectra usually agree to within a few percent in
the overlapping region around 1900--1950\,{\AA}). We remove 
foreground Galactic extinction using the dust maps of 
\citet*{1998ApJ...500..525S} and the Galactic extinction curve of
\citet*{1989ApJ...345..245C}. Furthermore, we co-add the spectra of 
galaxies
observed multiple times to increase the signal-to-noise ratio. For
the ultraviolet spectral slope to be measured with reasonable accuracy,
a minimum signal-to-noise ratio per pixel of 3 is required (see 
Appendix). A total of 110 galaxies with both far- and near-ultraviolet
spectra and 185 galaxies with only far-ultraviolet spectra satisfy this
cut in signal-to-noise ratio.

We require two additional conditions to include galaxies in our
final sample: that optical photometry be available to estimate
the ultraviolet aperture correction, and that {\it IRAS} 25, 60 and 
100\,$\mu$m observations be available to estimate the total luminosity
reradiated by dust. The above sample of 295 galaxies includes
the 57 starburst galaxies with optical angular diameters less than
$4\arcmin$ studied by \citet{1999ApJ...521...64M}, 
for which most of the 
concentrated ultraviolet emission could be observed within the
$20\arcsec \times 10\arcsec$ {\it IUE} aperture. {\it IRAS} 25, 60 and
100\,$\mu$m observations are available for 50 of the Meurer et al.
galaxies, which we include in our final sample. For other galaxies, 
reasonable ultraviolet aperture corrections may be derived as a 
function of the effective optical radius and morphological type 
(\citealt*{1995A&AS..114..527R}). Morphological types are known for
{\it IUE} 
galaxies, but effective optical radii must be obtained from independent
observations. We thus cross-correlate the above sample of 238 {\it IUE}
galaxies not included in the 
\citet{1999ApJ...521...64M} sample with the Data
Release One of the Sloan Digital Sky Survey (SDSS DR1; 
\citealt{2003AJ....126.2081A}).
In this way, we obtain effective $g$-band radii for 41 
galaxies (with typical, i.e. median $r_{\rm eff}\sim9\arcsec$), for 
which we compute ultraviolet aperture corrections using the 
prescription of \citet{1995A&AS..114..527R}. The corrections vary from 
galaxy to galaxy, with a mean logarithmic ratio of 
corrected-to-observed ultraviolet fluxes of $0.2 \pm0.2$. Only 33 of
these galaxies have available {\it IRAS} 25, 60 and 100\,$\mu$m 
observations and can be included in our final sample. These typically
have morphological types earlier than the starburst galaxies of the 
\citet{1999ApJ...521...64M} (Sa-Sc versus Sc-Im, although some 
starburst galaxies are also present) and similar absolute $B$-band 
magnitudes, $-16 \la M_B \la-22$ (we adopt a Hubble constant $H_0=
70\,$km\,s$^{-1}$\,Mpc$^{-1}$).

We supplement this sample with (non-Seyfert) galaxies for which 
multi-band ultraviolet imaging with {\it ANS} and {\it OAO} is available
from the compilation of \citet*{1995A&AS..109..341R}. In the 
spectral range of interest to us, the galaxies observed with {\it ANS}
were imaged at 1550, 1800, 2200 and 2500~{\AA} through a fixed aperture
of $2.5\arcmin\times2.5\arcmin$. Those observed with {\it OAO} were
imaged at 1550, 1910 and 2460~{\AA} through a fixed circular aperture
of $10\arcmin$ diameter. Of the 95 galaxies observed with {\it ANS} in
the \citet{1995A&AS..109..341R} compilation, 34 have available {\it IRAS} 
25, 60 and 100\,$\mu$m observations. We reject 5 galaxies with optical
diameters larger than $9\arcmin$, for which the ultraviolet emission 
does not appear to be clearly concentrated within the {\it ANS} 
aperture, and 2 galaxies with larger-aperture imaging available from {\it
OAO} (see below). The remaining 27 galaxies have typical morphological types
Sb-Sd and absolute $B$-band magnitudes similar to those of the galaxies
in the {\it IUE} sample, $-17 \la M_B \la-22$. Of the 36 galaxies with
{\it OAO} imaging, 11 have available {\it IRAS} 25, 60 and 100\,$\mu$m
observations.  We reject 6 galaxies with optical diameters larger than
$12\arcmin$. We thus retain only 5 galaxies with {\it OAO} imaging in 
our final sample, two Sc galaxies with $M_B\approx-16$, two Sd galaxies
with $M_B\approx-20$ and an Im galaxy with $M_B \approx-19$. Our final
sample contains 115 galaxies with both ultraviolet and infrared data, 
of which 83 were observed with {\it IUE}, 27 with {\it ANS} and 5 with
{\it OAO}. About half are starburst galaxies (most belonging to the 
\citealt{1999ApJ...521...64M} sample), while the other half are 
earlier-type, more 
quiescent galaxies spanning a similar range of absolute $B$-band
magnitudes.

We compute ultraviolet spectral slopes for all the galaxies in our
sample. \citet{1994ApJ...429..582C} define the ultraviolet spectral
slope $\beta$ by a power-law fit of the form $f_\lambda \propto 
\lambda^{\beta}$ (where $f_\lambda$ is the flux per unit wavelength)
to the spectrum in ten continuum bands in the range $1268\le \lambda
\le2580$~\AA. We cannot use this definition here, because it does
not apply to galaxies for which only multi-band ultraviolet imaging 
is available. Instead, we adopt the following definition of the 
ultraviolet spectral slope for all the galaxies in our sample,
\begin {equation}
\betaGLX = \frac{\log\,\ffuv - 
		   \log\,\fnuv}
		  {\log\,\lambda_{\rm FUV} - 
		   \log\,\lambda_{\rm NUV}},
\label{betadef}
\end {equation}
where $\lambda_{\rm FUV}=1520\,${\AA} and $\lambda_{\rm NUV}=
2310\,${\AA} are the effective wavelengths of the far- (1350--1800~\AA)
and near-ultraviolet (1800--3000~{\AA}) filters on board the {\it
Galaxy Evolution Explorer} ({\it GALEX},
\citealt{2003SPIE.4854..336M}), and \ffuv\ and \fnuv\
are the mean flux densities (per unit wavelength) through these filters.
In the Appendix, we use a reference sample of 110 {\it IUE} spectra to
calibrate the conversions between \betaGLX\ and the ultraviolet 
spectral slopes estimated from observations with different satellites.
We show that, in particular, ultraviolet spectral slopes estimated from
{\it ANS} and {\it OAO} imaging observations can be converted into 
\betaGLX\ with an rms uncertainty of only $\sigma \approx
0.2$. This is similar to the uncertainty introduced in \betaGLX\ by
the lack of near-ultraviolet spectra for some {\it IUE} galaxies in our
sample. For all the galaxies in our sample, we compute the ultraviolet
flux as $\Ffuv = \lambda_{\rm FUV} \ffuv$, as described in the Appendix.

To estimate the total far-infrared flux $F_{\rm dust}$ from the 
observed {\it IRAS} flux densities $f_{\nu} (25\,\micron)$, $f_{\nu}
(60\,\micron)$ and $f_{\nu}(100\,\micron)$, we appeal to the recent 
prescription of \citet{2002ApJ...576..159D}.
This is based on model fits to
the spectra of a large sample of nearby, normal (i.e.  non-Seyfert)
star-forming galaxies across the whole wavelength range from 3 to 
850\,$\mu$m. We use equation~(5) of Dale \& Helou to compute 
$F_{\rm dust}$ ($F_{\rm TIR}$ in their notation) from $f_{\nu} 
(25\,\micron)$, $f_{\nu} (60\,\micron)$ and $f_{\nu}(100\,\micron)$.
For reference, for the {\it IUE} starburst galaxies in our sample, 
the values of $F_{\rm dust}$ computed in this way are typically 50\% 
larger than those estimated by \citet{1999ApJ...521...64M} from only
$f_{\nu} (60\,\micron)$ and $f_{\nu}(100\, \micron)$ using the older 
prescription of Helou et al. (1988; see also 
\citealt{2000ApJ...533..682C}). Neglecting possible anisotropies 
(Section\,4), we equate the ratio of total far-infrared to ultraviolet
fluxes $F_{\rm dust}/\Ffuv$ to the corresponding luminosity ratio 
\ldustfuv.

\subsection{The observational \IRXUV\ diagram}

\begin{figure}
\includegraphics[height=85mm,angle=-90]{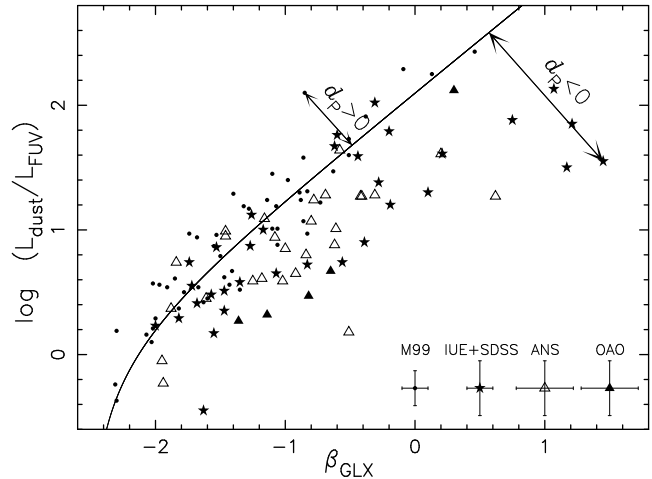}
\caption{
Ratio of far-infrared to ultraviolet luminosities plotted against
ultraviolet spectral slope (the `\IRXUV\ diagram'). The data points
are from the sample compiled in Section\,2.1 (median measurement 
errors for each subsample are indicated in the lower right corner).
The solid line shows the mean \IRXUV\ relation for the 50 starburst
galaxies in this sample (equation\,\ref{meansb}).
}
\label{irx}
\end{figure}

Fig.\,\ref{irx} shows \ldustfuv\ as a function of \betaGLX\ for the
115 galaxies in our sample. Different symbols distinguish the 50
starburst galaxies studied by Meurer et al. (1999, solid dots)
and the (typically more quiescent) galaxies observed with {\it 
IUE}+SDSS (stars), {\it ANS} (open triangles) and {\it OAO} (filled 
triangles). To a first, rough approximation, \ldustfuv\ is a function 
of \betaGLX\ for the whole sample. As found previously by Meurer et al.
(1995, 1999), the two quantities are tightly correlated for starburst
galaxies, for which the typical scatter at fixed \betaGLX\ amounts to
a factor of less than 2 in \ldustfuv. The main novelty introduced by
the inclusion of more quiescent galaxies in our sample is an increase
of the scatter to a factor of more than 3 toward lower \ldustfuv\ 
values at fixed \betaGLX.\footnote{It is worth noting that no galaxy
in Fig.~\ref{irx} lies substantially above the \IRXUV\ relation for
starburst galaxies. This is not unexpected, since our sample of galaxies
drawn from ultraviolet surveys (with typical infrared luminosities
$\ldust\sim10^{10}\,L_\odot$) does not include `luminous and 
ultraluminous infrared galaxies', which tend to have larger \ldustfuv\
than starburst galaxies at fixed \betaGLX\ (presumably because of the
heavy obscuration of the sources of ultraviolet radiation; 
\citealt{2002ApJ...568..651G}).} The effect is real, as the 
observational scatter is much larger than the typical (i.e.
median) measurement errors (which are shown in the lower right corner of 
Fig.\,\ref{irx}). We note that the {\em shape} of the \IRXUV\ relation
does not appear to be affected by the inclusion of quiescent galaxies
in our sample.

What are the main physical parameters driving the relation between 
ultraviolet spectral slope and ratio of far-infrared
to ultraviolet luminosities for star-forming galaxies? 
\citet{2000ApJ...539..718C}
show that, for starburst galaxies, the tight relation between
\betaGLX\ and \ldustfuv\ can be understood most simply as a sequence 
in the overall dust content of the galaxies. The inclusion of quiesent
star-forming galaxies in Fig.\,\ref{irx} reveals that a `second 
parameter' tends to broaden the relation.  We may begin by identifying
this second parameter as the perpendicular distance from the mean 
\IRXUV\ relation for starburst galaxies. Following 
\citet{1999ApJ...521...64M},
we define the mean relation for starburst galaxies by a 
least-squares fit of the type
\begin {equation}
L_{\rm dust}/L_{\rm FUV}
=
10^{\left(2.10 + 0.85 \betaGLX\right)} - 0.95
\label{meansb}
\end {equation}
to the observations of the 50 starburst galaxies in our sample. The
result is shown by the solid line in Fig.\,\ref{irx}. We can compute
the perpendicular (i.e. shortest) distance \dpp\ from each point in
the \IRXUV\ diagram to this line. We adopt the convention 
of negative \dpp\ for galaxies with \ldustfuv\ lower than the mean 
value for starburst galaxies at fixed \betaGLX\ (see Fig.\,\ref{irx}).

\begin{figure*}
\includegraphics[height=120mm,angle=-90]{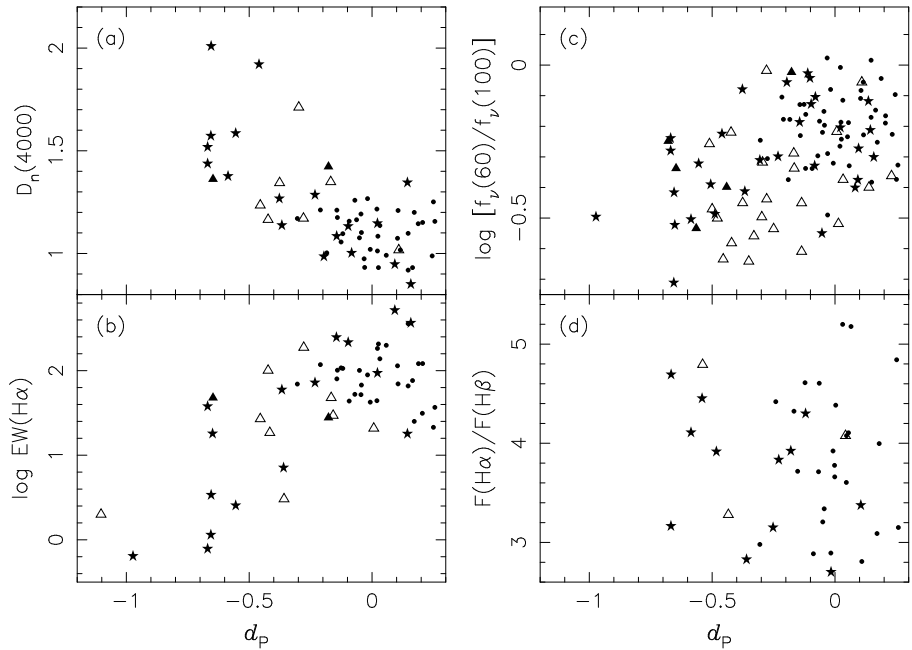}
\caption{
(a) spectral discontinuity at 4000\,{\AA}, (b) \ha\ emission 
equivalent width, (c) ratio of {\it IRAS} $f_{\nu} (60\,\micron)$
and $f_{\nu}(100\,\micron)$ flux densities and (d) \hahb\ ratio 
plotted as a function of perpendicular distance from the mean starburst
\IRXUV\ relation, for the subsets of galaxies for which these observed
quantities are available in the sample of Fig.\,\ref{irx}. The 
different symbols have the same meaning as in Fig.\,\ref{irx}.
}
\label{dpobs}
\end{figure*}

We expect the second parameter affecting the \IRXUV\ relation to be
connected to star formation history, since the relation becomes broader
when quiescent star-forming galaxies are considered in addition to 
starburst galaxies. It is thus natural to explore how observed 
quantities with different dependences on the star formation history 
correlate with \dpp. We first consider the 4000\,{\AA} spectral 
discontinuity index \dn,  defined as the ratio of the average flux 
densities in the narrow bands 3850--3950~{\AA} and 4000--4100~{\AA}
(\citealt{1999ApJ...527...54B}). This index depends somewhat on 
metallicity but correlates more with the ratio of present to 
past-averaged star formation rates in galaxies, and hence, it
is a valuable indicator of the history of star formation 
(\citealt{astro-ph/0311060}). Also, the narrowness of the bands used to
compute \dn\ makes it largely insensitive to attenuation by dust. 
Medium-resolution spectra (3--9\,{\AA} FWHM) of 64 galaxies in our 
sample (35 starburst and 29 quiescent star-forming) are available from
\citet{1992ApJS...79..255K}, \citet*{1995ApJS...97..331M},
\citet*{1995ApJS...98..103S}, \citet{2002A&A...389..845K} and
\citet{2003AJ....126.2081A}. Fig.\,\ref{dpobs}a shows \dn\ measured from
these spectra as a function of the perpendicular distance \dpp\ from the
mean starburst \IRXUV\ relation. The Spearman rank correlation 
coefficient is $r_s= -0.61$, indicating that the correlation between
\dn\ and \dpp\ is significant at the $5\,\sigma$ level for this sample
size. This result is remarkable, as it is direct observational evidence
that star formation history is relevant to the second parameter 
affecting the \IRXUV\ relation.

Another observable quantity that is sensitive to the ratio of 
present to past-averaged star formation rates is the \ha\ emission
equivalent width \ewha\ (\citealt{1983ApJ...272...54K};
\citealt*{1994ApJ...435...22K}). This is the ratio of the \ha\ 
emission-line luminosity produced by young massive stars to the 
continuum luminosity per unit wavelength produced by older stars at 
$\lambda_{\rm H\alpha}=6563$\,\AA. In Fig.\,\ref{dpobs}b, we show \ewha\
as a function of \dpp\ for 59 galaxies (31 starburst and 28 quiescent
star-forming) in our sample (we adopt the convention of positive 
equivalent width for emission). The \ha\ emission equivalent widths were 
measured from the same optical spectra as above following the procedure
outlined in section\,2 of \citet{astro-ph/0311060}, which allows accurate 
correction for stellar absorption. The Spearman rank correlation 
coefficient in this case is $r_s= 0.48$, indicating that the correlation
between \ewha\ and \dpp\ is significant at the $4\,\sigma$ level for the
sample size. We note that dust can reduce
\ewha\ through the absorption of ionizing photons.
However, the effect is modest for quiescent star-forming galaxies, 
typically only 20--30 per cent, and cannot account for the trend
seen in Fig.\,\ref{dpobs}b (\citealt{2002MNRAS.330..876C}). Taken
together, Figs\,\ref{dpobs}a and b show that two different tracers
of the ratio of present to 
past-averaged star formation rates, \dn\ and \ewha,
correlate with \dpp. This is strong evidence that the ratio of present
to past-averaged star formation rates could be the second parameter 
affecting the \IRXUV\ relation.

In an attempt to further understand the relative influence of star
formation history and attenuation by dust on the \IRXUV\ relation,
we plot other properties of the galaxies against \dpp\ in 
Figs\,\ref{dpobs}c and d: the ratio of the {\it IRAS} $f_{\nu}
(60\,\micron)$ and $f_{\nu}(100\,\micron)$ flux densities and the
ratio of the \ha\ and \hb\ emission line luminosities. The \f60100\
ratio, available for all 115 galaxies in our sample, depends on
the relative contributions by young and old stars to dust heating in a 
galaxy (e.g., \citealt{1986ApJ...311L..33H}).
Hence it is an indirect tracer of the star
formation history that also depends on the dust content (and 
potentially the spatial distribution of dust relative to stars). The
\hahb\ ratio, available for 48 galaxies (29 starburst and 19 quiescent 
star-forming) from the same optical spectra as above, is strongly
sensitive to attenuation by dust but not to the past history of star
formation. The Spearman rank correlation coefficients are $r_s=+0.43$
for the relation between \f60100\ and \dpp\ and $r_s=0.12$ for that
between \lhahb\ and \dpp. For the different numbers of galaxies in 
each case, this implies that the correlation between \f60100\ and 
\dpp\ is significant at the $4\,\sigma$ level, while there appears
to be little or no correlation between \lhahb\ and \dpp. These 
findings reinforce our contention that the second parameter affecting
the \IRXUV\ relation is primarily star formation history rather than
dust content.

Based on the observational evidence above, we 
conclude that, while dust content is the primary driver of the \IRXUV\
relation for starburst galaxies, the broadening of the relation for
quiescent star-forming galaxies is most likely driven by star formation
history, and in particular the ratio of present to past-averaged star
formation rates. In the next section, we show that this observational
result can be understood in the framework of a simple model for the
production of starlight and its transfer through the interstellar 
medium (ISM) of galaxies.

\section{The origin of the \IRXUV\ relation: Modelling}

In this section, we explore the physical origin of the \IRXUV\ relation
from a more theoretical standpoint. We appeal to a simple model for 
the production of starlight and its transfer through the interstellar
medium in galaxies. The model is based on a combination of the 
\citet{2003MNRAS.344.1000B} population synthesis code and the 
\citet{2000ApJ...539..718C}
prescription for the absorption of starlight by dust. The
Charlot \& Fall model relies on an idealized description of the main 
features of the ISM. Stars are assumed to form in interstellar `birth 
clouds' (i.e.  giant molecular clouds). After $10^7\,$yr, young stars 
disrupt their birth clouds and migrate into the `ambient ISM'. The 
`effective absorption' curve describing the attenuation of photons 
emitted in all directions by stars of age $t'$ in a galaxy is given by
the simple formula
\begin{eqnarray}
\hat{\tau}_\lambda(t')=\cases{
\hskip0.22cm\hat{\tau}_V\left(\lambda/{5500\,\mathrm{\AA}}\right)^{-0.7}\,,
&for $t'\leq 10^7$ yr,\cr
{{\mu\hat{\tau}_V}}\left(\lambda/{5500\,\mathrm{\AA}}\right)^{-0.7}\,,
&for $t'>10^7$ yr,\cr}
\label{taueff}
\end{eqnarray}
where $\hat{\tau}_V$ is the total effective $V$-band optical depth seen
by young stars. The adjustable parameter $\mu$ defines the fraction of 
the total effective absorption optical depth contributed by the ambient
interstellar medium ($\mu \approx1/3$ on average, with substantial 
scatter). In addition, a fraction $f\approx0.1$ of the effective 
absorption optical depth $(1-\mu)\hat{\tau}_V$ of the birth clouds
is assumed to arise from the \hii\ regions ionized by young stars 
in the inner parts of these clouds. We note that, in the case of 
optically thick birth clouds, the ultraviolet radiation escaping from
a galaxy is produced by stars older than $\sim10^7\,$yr, while the total
far-infrared luminosity is produced by stars of all ages. 

\citet{2000ApJ...539..718C} used their model to interpret the relations
between ultraviolet spectral slope, ratio of far-infrared to 
ultraviolet luminosities and \hahb\ ratio for the starburst galaxies 
in the \citet{1999ApJ...521...64M} 
sample (Section~2). They showed that the
finite lifetime of the birth clouds is a key ingredient for resolving
the apparent discrepancy between the attenuation of line and continuum
radiation in these galaxies. They also showed that the \IRXUV\ relation
can be understood most simply as a sequence in the overall dust content
of the galaxies. For simplicity, \citet{2000ApJ...539..718C} limited 
their study to models with constant star formation rates and young 
ages, as appropriate to interpret observations of starburst galaxies.
The results of Section\,2 above suggest that star formation history 
plays an important role in defining the \IRXUV\ relation for more 
quiescent galaxies. To investigate the dependence of the \IRXUV\ 
relation on star formation history, we assume that the star formation
rate varies with time as
\begin{equation}
\psi(t)\propto\exp(-\gamma\,t)
\label{sfr}
\end{equation}
from $t=0$ to the present galaxy age $t=\tp$. The parameter $\gamma$
is the inverse time-scale of star formation. For simplicity, all 
models are assumed for the moment to have fixed solar metallicity.

Our main goal is to identify the model parameter that most simply 
accounts for the offset of normal, quiescent star-forming galaxies 
from the mean \IRXUV\ relation for starburst galaxies in 
Fig.\,\ref{irx}. To do this, we must explore the influence of each
primary model parameter (\tp, $\gamma$, \tauV\ and $\mu$) on
observable quantities available for galaxies in our sample, i.e. 
the ultraviolet spectral slope, the ratio of far-infrared to 
ultraviolet luminosities and the 4000\,{\AA} discontinuity.
The influence of \tauV\ and $\mu$ on \betaGLX\ and \ldustfuv\ can be
inferred from the results presented in figure\,2 of 
\citet{2000ApJ...539..718C}:
these parameters, in particular the effective dust absorption
optical depth in the ambient ISM, $\mu\tauV$, control the position of 
a galaxy {\em along} the \IRXUV\ relation. 
The narrow-band spectral index \dn\ is only weakly sensitive to 
variations in \tauV\ and $\mu$ (Section\,2). We find that the influence
of the star formation history on the \IRXUV\ relation through the 
parameters \tp\ and $\gamma$ is best described in terms of the 
ratio of present to past-averaged star formation rates. This is defined
by
\begin{equation}
b=\frac{
	  \psi(\tp)}
	  {
	  \langle\psi\rangle}
    =\frac{
	  \tp\,\psi(\tp)}
	  {
	  \int_0^{\tp}\,dt'\,\psi(t')}\,.
\label{bparam}
\end{equation}
For the expression of $\psi(t)$ in equation~(\ref{sfr}), we have
\begin{equation}
b=\frac{
	\gamma\tp\,
	\exp\left(-\gamma\tp\right)}
	 {
	1-\exp\left(-\gamma\tp\right)}\,.
\label{b2param}
\end{equation}
For reference, a galaxy with constant star formation rate ($\gamma=0$)
has $b=1$ at all ages.\footnote{Since the age \tp\ of the oldest stars
in a galaxy is difficult to estimate observationally, it is often 
replaced by some reference age, for example the age of the Universe 
(e.g. \citealt{1983ApJ...272...54K};
\citealt*{1994ApJ...435...22K}). Such an 
approximation leads to larger $b$ than obtained when using the 
definition given in equation~(\ref{bparam}).}

\begin{figure}
\includegraphics[width=100mm,angle=-0]{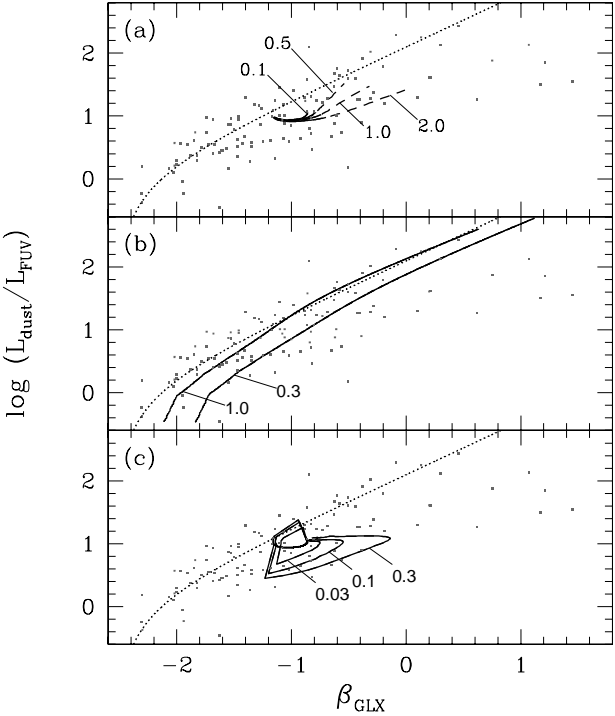}
\caption{(a) Age sequences of models with star formation histories given
by equation\,(\ref{sfr}) with $\gamma=0.1$, 0.5, 1.0 and 2.0\,Gyr$^{-1}$
(as indicated) and for $\tauV=1.5$ and $\mu=0.3$. The sequences are
plotted as solid lines at ages when the ratio of present to past-averaged
star formation rates is larger than $b=0.3$ and as dashed lines at
greater ages, down to $b=0.03$. (b) Sequences of models with different 
dust content (i.e. different \tauV) and $\mu=0.3$, for two ratios of 
present to past-averaged star formation rates, $b=1.0$ and 0.3 (as 
indicated). (c) Age sequences of models with $\gamma=0.15$ (equation 
\ref{sfr}), which undergo an extra burst of star formation of duration 
$10^8\,$yr at the age of 6\,Gyr involving 3, 10 and 30 per cent of the 
total mass of stars formed by the continuous model over 13.5\,Gyr (as 
indicated). The dust parameters are $\tauV=1.5$ and $\mu=0.3$. In all 
panels, the data and the mean starburst \IRXUV\ relation shown as a 
dotted line are the same as in Fig.~\ref{irx}.}
\label{irxbpar}
\end{figure}

Fig.\,\ref{irxbpar} summarizes the main results of our analysis.
In Fig.\,\ref{irxbpar}a, we show \ldustfuv\ as a function of \betaGLX\
for sequences of models with different ages for four different 
time-scales of star formation, corresponding to $\gamma=0.1$, 0.5, 1.0 
and 2.0\,Gyr$^{-1}$, and for $\tauV=1.5$ and $\mu=0.3$. The sequences are 
plotted as solid lines at ages when the ratio of present to past-averaged
star formation rates is larger than $b=0.3$ and as dashed lines at 
greater ages, down to $b=0.03$. From $b=1$ (age $t=0$) to $b=0.3$, 
the distance from the mean 
starburst relation increases as $b$ decreases along the model sequences,
models with $b\approx0.3$ occupying regions of the diagram populated
by more quiescent galaxies.  This behavior arises mainly from an 
increase in \betaGLX\ as $b$ decreases. Models with $b \approx0.3$
are dominated by stellar populations with intrinsically
redder spectra and hence larger ultraviolet spectral slopes than 
starburst galaxies. We note that, for $b>0.3$, all models in 
Fig.\,\ref{irxbpar}a follow a similar path in the \IRXUV\ diagram. For
$b<0.3$, the evolution in the \IRXUV\ diagram depends more strongly on
the adopted time-scale of star formation. At fixed $b$, \betaGLX\ is 
larger in models with short time-scales of star formation (large 
$\gamma$), which contain a smaller proportion of ultraviolet-bright 
stars than models with extended star formation. In this $b$ range, 
\ldustfuv\ increases at late ages because of the contribution to dust
heating by the optical radiation from old stars (Section\,4). We note
that models with small $b$ `returning' at late ages to the mean \IRXUV\
relation for starburst galaxies are considerably fainter than starburst
galaxies and have intrinsically red optical spectra.

The similarity for $b\ga0.3$ between models with different $\gamma$ in
the \IRXUV\ diagram prompts us to plot, in Fig.\,\ref{irxbpar}b,
\ldustfuv\ as a function of \betaGLX\ for sequences of models with 
different dust content (i.e. different \tauV), for two ratios of present
to past-averaged star formation rates, $b=1.0$ and 0.3. For simplicity, 
we assume that the fraction of the total effective absorption optical 
depth contributed by the ambient ISM is fixed at the standard value 
$\mu=0.3$ in all models.  The striking result of Fig.\,\ref{irxbpar}b is
that the model sequences with constant $b$ are remarkably parallel to the
mean \IRXUV\ relation for starburst galaxies, the $b=0.3$ sequence
being offset toward the region of the diagram populated by quiescent 
star-forming galaxies. This is consistent with
the observational evidence from Figs\,\ref{dpobs}a and b that, while
dust content is the primary driver of the \IRXUV\ relation, the 
broadening of the relation for quiescent star-forming galaxies is most
likely driven by star formation history. This result does not depend on
the specific combination of $\gamma$ and $\tp$ used to compute models
with $b=0.3$ in Fig.\,\ref{irxbpar}b. We find that, for $b\ll0.3$, 
sequences of models with different dust content are also parallel to 
the mean \IRXUV\ relation for starburst galaxies. However, as expected
from Fig.\,\ref{irxbpar}a, the location of the sequences in this $b$\
range depends sensitively on the choices of $\gamma$ and $\tp$ (see 
also Section\,4).

Figs\,\ref{irxbpar}a and b suggest that models with simple continuous 
star formation histories cannot easily reproduce the observations of 
galaxies lying at the largest perpendicular distances from the mean 
starburst \IRXUV\ relation in our sample, unless very short time-scales
of star formation ($\gamma\sim2$) and very small ratios of present to
past-averaged star formation rates ($b\sim0.03$) are invoked. Such 
parameters are not typical of quiescent star-forming galaxies. In 
Fig.\,\ref{irxbpar}c, we show \ldustfuv\ as a function of \betaGLX\ 
for models with a star formation time-scale parameter $\gamma=0.15$,
more typical of a spiral galaxy, which undergo an extra burst of star 
formation of duration $10^8\,$yr at the age of 6\,Gyr, for $\tauV=1.5$
and $\mu=0.3$. We show the time evolution of these models for bursts 
representing 3, 10 and 30 per cent of the total mass of stars formed by
the continuous model over 13.5\,Gyr. When the burst starts, each model 
first joins the mean \IRXUV\ relation for starburst galaxies. Shortly
after the burst ends, \ldust\ and hence \ldustfuv\ drop rapidly because
of the dispersal of the (optically thick) birth clouds of the last 
stars formed during the burst. Then, \betaGLX\ and \ldustfuv\ increase
again as the ultraviolet light from these stars fades and reddens. A
maximum in \betaGLX\ is reached about 0.5--1\,Gyr after the burst ended.
The subsequent evolution toward smaller \betaGLX, as the underlying 
continuous star formation again starts to dominate the production of the
ultraviolet light, is remarkable for two reasons. First, it is very slow,
implying that the models spend several gigayears in the region of the 
diagram populated by galaxies far away from the mean starburst \IRXUV\ 
relation. Second, the $b$ parameter is only slightly lower than that of
the undisturbed continuous model and hence typical of quiescent 
star-forming galaxies (note that this conserves the trend of decreasing
$b$ with increasing distance from the mean starburst relation). Thus,
models including minor bursts of star formation on top of continuous star
formation histories can account in a natural way for the observations of
galaxies at large distances from the mean starburst relation in the
\IRXUV\ diagram.

The simple model described above provides a natural framework for 
understanding the physical origin of the \IRXUV\ diagram. As shown in
Fig.\,\ref{irxbpar}, this model can account for the observational 
evidence from Section\,2 that, while dust content is the primary driver
of the \IRXUV\ relation for starburst galaxies, the broadening of the
relation for quiescent star-forming galaxies is most likely driven by
star formation history, and in particular the $b$ parameter. These 
results do not depend sensitively on the population synthesis code 
used for computing the production of starlight in the model. We have
checked that adopting the {\small P\'EGASE} population synthesis code 
(\citealt{1997A&A...326..950F}; version 2.0) instead of the 
\citet{2003MNRAS.344.1000B} code has a negligible influence on the 
results of Fig.\,\ref{irxbpar}.

It is worth noting that other effects, such as anisotropic emission from
galaxies and spatial distribution of the dust, can potentially affect
the positions of galaxies in the \IRXUV\ diagram. The inclination angles
available for 59 quiescent star-forming galaxies observed with {\it 
IUE}+SDSS, {\it ANS} and {\it OAO} in our sample span the full range from
0 to 90 degrees and do not exhibit any specific trend with perpendicular 
distance from the mean \IRXUV\ starburst relation. Therefore, we do not 
expect orientation effects to play a major role in the broadening of the
\IRXUV\ relation for these galaxies. \citet{2000ApJ...539..718C} show 
that, for some combinations of optical properties and spatial 
distributions of the dust in their model, it is possible to produce
\ldustfuv\ ratios lower than observed for starburst galaxies at large
ultraviolet spectral slopes. However, the results in their figures 6
and 13 indicate that such models could not account for the low \ldustfuv\
ratios of quiescent star-forming galaxies with small \betaGLX\ in our 
sample.\footnote{\citet{2000ApJ...528..799W} also explore the 
influence of the optical properties and spatial distribution of the dust
on the \IRXUV\ relation using a code of radiative transfer with spherical
symmetry. None of the models they investigate produces \ldustfuv\ ratios
lower than observed for starburst galaxies at fixed ultraviolet spectral
slope (see their figure 12, mislabeled figure 13 in the print version).}
Changes in the spatial distribution of the dust also do not provide a
natural explanation for the strong drop in \ha\ equivalent width with
increasing distance from the mean starburst \IRXUV\ relation in
Fig.\,\ref{dpobs}b (see figure 6c of \citealt{2000ApJ...539..718C}). We
conclude that, although orientation and dust geometry may have minor
effects on the positions of galaxies in the \IRXUV\ diagram, they do not
seem responsible for the overall broadening of the \IRXUV\ relation for 
quiescent star-forming galaxies in our sample. This broadening can be 
most simply understood as a sequence in the star formation histories of
the galaxies.

\section{Constraints on the ultraviolet attenuation}

The results presented in the previous sections have important 
implications for estimates of the ultraviolet attenuation \Afuv\ and
hence of star formation rates in galaxies. \citet{1999ApJ...521...64M}
derive 
simple formulae for estimating \Afuv\ from either the ultraviolet 
spectral slope or the ratio of far-infrared to ultraviolet 
luminosities on the basis of the tight \IRXUV\ relation for starburst
galaxies.  The results of Sections\,2 and 3 above suggest that 
these formulae cannot be straightforwardly applied to more quiescent 
galaxies. We can use the model described in Section\,3 to estimate
the accuracy to which \Afuv\ can be constrained in galaxies of 
different types. A way to assess this is to investigate the relations
between \Afuv, \betaGLX\ and \ldustfuv\ for a library of models 
encompassing a wide range of physically plausible star formation 
histories and dust contents. We thus generate a library of Monte 
Carlo realizations of different star formation histories similar
to that used by \citet{2003MNRAS.341...33K} 
to interpret the \dn\ and \hda\
index strengths of a complete sample of over $10^5$ SDSS galaxies. 

The main requirement for the model library is to include broad enough
ranges of star formation histories and dust contents to ensure a wide 
coverage of the \IRXUV\ diagram. We have checked that the results 
presented below do not depend sensitively on the detailed parametrization
of the models. For simplicity, we follow \citet{2003MNRAS.341...33K} and 
parametrize each star formation history in terms of two components: an
underlying continuous model with a star formation law given by 
equation~(\ref{sfr}), and random bursts superimposed on this continuous
model.\footnote{Specifically, we take the galaxy age \tp\ to be 
distributed uniformly over the interval from 1.5 to 13.5\,Gyr
and the star formation time-scale parameter $\gamma$
over the interval from 0 to 1~Gyr$^{-1}$. Random bursts occur with equal
probability at all times until \tp. They are parametrized in terms of
the ratio between the mass of stars formed in the burst and the total
mass of stars formed by the continuous model over the time \tp. The 
ratio is taken to be distributed logarithmically from 0.03 to 4.0. During
a burst, stars form at a constant rate for a time distributed uniformly
in the range $3\times 10^7$--$3 \times 10^{8}$ years. The burst 
probability is set so that 50 per cent of the galaxies in the library
have experienced a burst in the past 2 Gyr (see 
\citealt{2003MNRAS.341...33K}).}
For the purpose of the present analysis, we retain only models with 
ratios of present to past-averaged star formation rates larger than 
0.03. For the attenuation by dust, we adopt a broad distribution of 
\tauV\ with a peak around 1.5 and a broad distribution in $\mu$ with a 
peak around 0.3. This is consistent with the dust properties derived 
from the \hahb\ ratios and broad-band optical magnitudes of the above 
sample of $10^5$ SDSS galaxies (\citealt{astro-ph/0311060}; 
\citealt{2003MNRAS.341...33K}).
We distribute our models logarithmically in metallicity from 0.1 to 2 
times solar. Our final library consists of 95,000 different models.
 
\begin{figure}
\includegraphics[width=85mm,angle=-0]{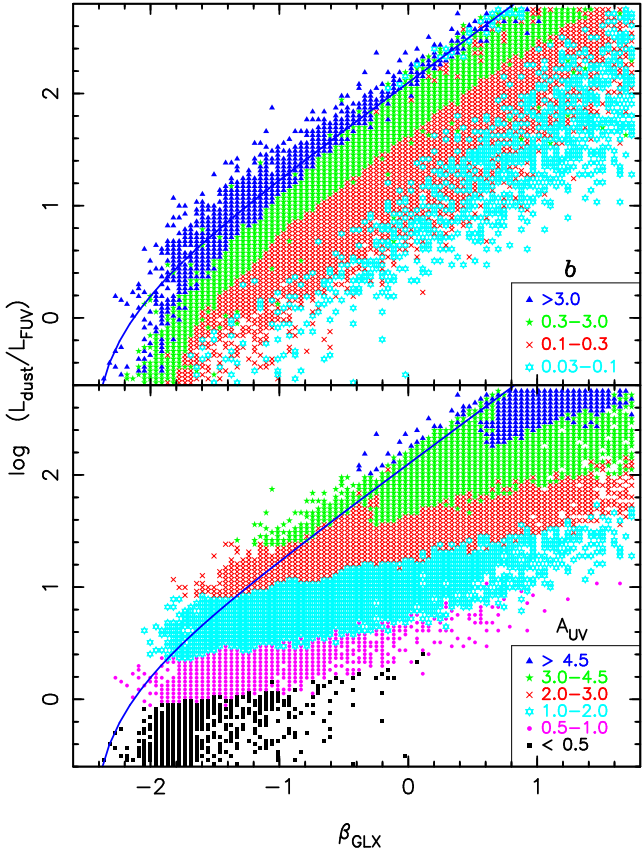}
\caption{Ratio of far-infrared to ultraviolet luminosities plotted
against ultraviolet spectral slope (the `\IRXUV\ diagram') for the
model library described in Section\,4. Upper panel: the \IRXUV\
plane has been binned and coded to reflect the average ratio of 
present to past-averaged star formation rates of the galaxies that
fall into each bin. Lower panel: the \IRXUV\ plane has been binned 
and coded to reflect the average ultraviolet attenuation \Afuv\ of
the galaxies that fall into each bin. In each panel, the solid line
shows the mean starburst \IRXUV\ relation from Fig.~\ref{irx}.}
\label{modelib}
\end{figure}

Fig.\,\ref{modelib} shows \ldustfuv\ as a function of \betaGLX\ for
these models. In the upper panel, we have divided the \IRXUV\ plane
into bins and coded each bin with a different colour to reflect the
average ratio of present to past-averaged star formation rates of
the simulated galaxies that fall into it. As found in Section\,3 above,
sequences of galaxies with different dust contents corresponding
to different $b$ lie parallel to the mean \IRXUV\ relation for
starburst galaxies. In the lower panel of Fig.\,\ref{modelib}, we
have coded each bin in the \IRXUV\ plane with a different colour to
reflect the average ultraviolet attenuation \Afuv\ for the galaxies
that fall into it. It is remarkable that sequences of models 
with different star formation histories corresponding to different
\Afuv\ are nearly perpendicular to the $y$-axis in this diagram.
This implies that \ldustfuv\ is a much better estimator of \Afuv\ than
\betaGLX\ in galaxies with different star formation histories. 

To quantify the accuracy to which \Afuv\ can be estimated from either
\ldustfuv\ or \betaGLX, we have explored functional fits of \Afuv\
in terms of these parameters using the models in our library. We find
that the dependence of \Afuv\ on \ldustfuv\ is well approximated by
two linear relations with a change in slope at some characteristic
turn-around point (corresponding to $\Afuv\approx1$). This can be 
represented by an expression of the form
\begin{eqnarray}
\Afuv&=&\frac{(A_0+A_1)(x-x_0)}{2} +\cr
     &+& \frac{ (A_1-A_0)\sigma_0}{2} \ln \left[\cosh
  	\left(\frac{x-x_0}{\sigma_0}\right)\right] + A_2\,,
\label{eq:irxfit}
\end{eqnarray}
where $x\equiv\log(\ldustfuv)$ and $A_0$ (the slope of the relation at small
$x$), $A_1$ (the slope of the relation at large $x$), $A_2$ (the 
normalization constant), $x_0$ (the turn-around point) and $\sigma_0$
(the smoothness of turn-around) are fitted parameters. The accuracy of 
the \Afuv\ estimates obtained from this formula depends on whether 
or not independent information is available on the ratio of 
present to past-averaged star formation rates. If no information is 
available on $b$, equation~(\ref{eq:irxfit}) with $A_0=0.39$, $A_1=
1.86$, $A_2=0.75$, $x_0=0.26$ and $\sigma_0=0.42$ provides estimates
of \Afuv\ with a typical (i.e. rms) relative uncertainty of 20 per cent
for all the models in the library.

\begin{table*}
\centering
\begin{minipage}{121mm}
\caption{Parameters of the fits of equation~(\ref{eq:irxfit}) for 
different ranges of present to past-averaged star formation rates
$b$. Also indicated for reference are the corresponding ranges in 
specific star formation rate $\psi(\tp)/\mstar$ (a quantity more
loosely connected to the star formation history but easier
to constrain observationally). The last row shows the results when 
no restriction is applied on $b$ and hence $\psi(\tp)/\mstar$. Also
listed is the typical (i.e. rms) relative uncertainty \srel\ in the
estimated \Afuv\ for the library of models described in Section\,4.}
\label{tab:irxfit}
\begin{tabular}{@{}llccccccc@{}}
\hline
Range in $\log\,b$ &
Range in $\log\left\{\left[\psi(\tp)/\mstar\right]({\rm yr}^{-1})\right\}$
& $A_0$ & $A_1$ & $A_2$ & $x_0$ & $\sigma_0$ & \srel\
\\
\hline
$[-1.5\,,\,-1.0]$  & 
$[-11.3\,,\,-10.6]$
& 0.27 &   2.02 &   1.01 &   0.69 &   0.91 &  0.15\\
$[-1.0\,,\,-0.5]$  & 
$[-10.6\,,\,-10.0]$
& 0.34 &   2.07 &   0.90 &   0.44 &   0.62 &  0.10\\
$[-0.5\,,\,+0.0]$  & 
$[-10.0\,,\,\,\,\,-9.3]$
& 0.35 &   2.24 &   0.92 &   0.36 &   0.60 &  0.06\\
$[+0.0\,,\,+2.0]$  & 
$[\,\,\,-9.3\,,\,\,\,\,-7.5]$
& 0.39 &   2.37 &   0.96 &   0.34 &   0.59 &  0.09\\
Full range         & 
Full range
& 0.39 &   1.86 &   0.75 &   0.26 &   0.42 &  0.19\\
\hline
\end{tabular}
\end{minipage}
\end{table*}

More accurate estimates of \Afuv\ can be obtained when even rough 
information is available on $b$. This is illustrated in 
Table\,\ref{tab:irxfit}, in which we list the parameters of the fits
and the accuracy of the \Afuv\ estimates obtained from 
equation~(\ref{eq:irxfit}) for different ranges in $\log\,b$. For
reference, we also indicate the corresponding ranges in `specific star
formation rate' $\psi(\tp)/\mstar$ (i.e. the current star formation 
rate per unit stellar mass), a quantity more loosely connected to the
star formation history but easier to constrain observationally. The typical
relative uncertainty in the \Afuv\ estimates decreases from 15 per cent for 
the most quiescent galaxies ($\log\,b <-1$) to only 7 per cent for 
starburst galaxies ($\log\,b>0$). The reason for this is that, in 
starburst galaxies, the heating of the dust and the production of the
ultraviolet radiation are dominated by the same stars. Thus \ldustfuv\
is intimately related to \Afuv. In quiescent galaxies, \ldustfuv\ 
is affected by dust heating by the optical radiation from old stars
with little ultraviolet emission, which depends on the history of
star formation at fixed \Afuv. For reference, the contribution to dust 
heating by stars older than 0.3\,Gyr ranges from typically less than 10
per cent for $b\ga1$ to about 80 per cent for $b\approx0.03$ in the 
models of Fig.\,\ref{modelib}. This is the reason why, when using the
fitting formulae derived from Table\,\ref{tab:irxfit}, the same \Afuv\ is 
obtained for larger \ldustfuv\ when $b$ is lower. A similar result was
noted by \citet{1999A&A...352..371B} 
in their analysis of a far-infrared selected
sample of nearby star-forming galaxies. We conclude that, if rough 
constraints are available on the ratio of present to past-averaged star
formation rates, reliable estimates of \Afuv\ can
be derived from \ldustfuv. We note that, for the models in our library,
$\log\,b$ can be roughly estimated from \dn\ as
\begin{equation}
\log\,b=\,-2.52\,+\,40.81\,\exp\left[-2.24\,\dn\right]\,,
\label{logbdn}
\end{equation}
with a dispersion of about 0.3 at fixed \dn. The corresponding dependence
of the specific star formation rate $\psi(\tp)/\mstar$ on \dn\ is in 
excellent agreement with that found by \citet{astro-ph/0311060} for the
sample of $10^5$ SDSS galaxies mentioned above.

In some cases, the far-infrared luminosity may not be available to 
estimate \Afuv\ (for example, for high-redshift galaxies for which 
only the rest-frame ultraviolet emission can be observed). In such 
cases, it is important to know the accuracy to which \Afuv\ can be 
estimated based solely on the ultraviolet spectral slope. The wide 
range in \Afuv\ spanned by models with fixed \betaGLX\ in 
Fig.\,\ref{modelib} indicates that \betaGLX\ is not a precise estimator
of \Afuv\ in galaxies with different star formation histories. This
is consistent with the results of \citet{2002ApJ...577..150B}. In 
fact, we did not succeed in finding a 
function of \betaGLX\ that provided reasonably accurate estimates of
\Afuv\ for the models in our library. Significant improvement is
obtained when including the ratio of present to past-averaged star
formation rates as a variable of the fit. In particular, we find that
the dependence of \Afuv\ on \betaGLX\ and $b$ is reasonably well 
approximated by
\begin{equation}
\Afuv= 3.87 + 1.87\left(\betaGLX\,+\,0.40\,\log\,b\right)\,.
\label{eq:betafit}
\end{equation}
This formula provides estimates of \Afuv\ for which the accuracy is an
increasing function of $b$. For $b\ga0.3$ (corresponding roughly to types 
later than Sb), the typical (i.e. rms) absolute uncertainty in the
\Afuv\ estimates is only 0.32. For lower $b$, the uncertainty is about
one magnitude. The reason for the large uncertainty in
\Afuv\ at low $b$ is that, in galaxies with little current star formation,
the ultraviolet spectral slope depends strongly on both dust and the 
history of star formation. We note that, for $b\approx5$, the expression
of \Afuv\ given in equation~(\ref{eq:betafit}) is similar to the
non-$b$-dependent expression for \Afuv\ as a function of ultraviolet
spectral slope
derived by Meurer et al. (1999, their equation\,11) from the analysis
of an ultraviolet-selected sample of nearby starburst galaxies. In the
absence of far-infrared observations, therefore, \Afuv\ can still be 
reasonably well estimated from the ultraviolet spectral slope alone, 
provided that some constraints are available on the ratio of 
present to past-averaged star formation rates.

\section{Summary and conclusion}

We have compiled a sample of 115 nearby, non-Seyfert galaxies drawn
from ultraviolet surveys spanning a wide range of star formation 
activities, from starburst to nearly dormant. Based on this
sample, we showed that, at fixed ultraviolet spectral slope (\betaGLX),
normal star-forming galaxies have systematically lower ratio of total
far-infrared to ultraviolet luminosities (\ldustfuv) than starburst 
galaxies. The tightness of the correlation between \betaGLX\ and 
\ldustfuv, shown by \citet{1999ApJ...521...64M} to provide valuable 
constraints on the ultraviolet attenuation in starburst galaxies, 
does not extend therefore to normal, more quiescent star-forming
galaxies. We showed that, for the galaxies in our sample, the strengths
of the age-sensitive 4000\,{\AA} spectral discontinuity and
\ha\ emission equivalent width correlate well
with distance from the mean relation for starburst galaxies in the 
\IRXUV\ diagram, while there is little or no correlation between the
dust-sensitive \hahb\ ratio and this distance. This is strong
observational evidence that, while dust content is the primary driver
of the \IRXUV\ relation for starburst galaxies, the broadening of the
relation for quiescent star-forming galaxies is most likely driven by
star formation history.

We have used the simple but physically realistic attenuation model of
\citet{2000ApJ...539..718C}, in combination with the new 
\citet{2003MNRAS.344.1000B}
population synthesis code, to investigate the physical origin
of the \IRXUV\ relation for galaxies with different star formation
histories. We confirm that, for starburst galaxies, the tight \IRXUV\
relation can be understood most simply as a sequence in overall dust
content. In addition, within the framework of this model, the second 
dimension of the \IRXUV\ relation can be understood most simply as a
sequence in the ratio of present to past-averaged star formation 
rates. 

The dependence of the \IRXUV\ relation on both star formation history
and dust content implies that the ultraviolet attenuation \Afuv\ cannot
be derived from \ldustfuv\ and \betaGLX\ as straightforwardly as expected
from the analysis of starburst galaxies alone. We have used a library of
Monte Carlo realizations of galaxies with different star formation 
histories and dust contents to quantify the accuracy to which \Afuv\ 
can be estimated from either \ldustfuv\ or \betaGLX. We find that 
\ldustfuv\ is a better estimator of \Afuv\ than \betaGLX\ in galaxies 
with different star formation histories. We provide simple formulae for 
estimating \Afuv\ as a function of either \ldustfuv\ or \betaGLX\ and
show that the accuracy of these estimates can be significantly improved
if some constraints are available on the ratio of present to 
past-averaged star formation rates (for example, based on the observed 
4000\,{\AA} discontinuity). The \IRXUV\ diagram, therefore, is a valuable
diagnostic of the attenuation by dust and hence the star formation rates
of galaxies in a wide range of star formation activities. This is 
especially important for applications to studies of high-redshift 
galaxies, for which the rest-frame optical and infrared emission cannot
be observed at present.

\section*{Acknowledgments}
X.K., S.C. and J.B. thank the Alexander von Humboldt Foundation, the
Federal Ministry of Education and Research, and the Programme for 
Investment in the Future (ZIP) of the German Government for financial
support. We are grateful to the referee, Eric Bell, for helpful comments.
This work made use of the HyperLeda astronomy information system 
(http://leda.univ-lyon1.fr/).

\appendix

\section{Calibration of ultraviolet spectral slopes and
fluxes}

Here, we use the sample of 110 high-signal-to-noise {\it IUE} galaxies
with both far- and near-ultraviolet spectra compiled in Section\,2.1
to calibrate the conversions between \betaGLX\ 
(equation\,\ref{betadef}) and the ultraviolet spectral slopes estimated
from observations with different satellites. The minimum 
signal-to-noise ratio per pixel\footnote{This
is the average signal-to-noise ratio per pixel in 100\,{\AA}-wide
continuum windows free of emission features centred at rest wavelengths
1450, 1700, 2300 and 2700~\AA\ (\citealt{1991ApJS...75..645K}).}
of 3 required for this sample ensures
that \betaGLX\ can be measured to an accuracy of better than 0.3.
For reference, the
median error in \betaGLX\ for the 110 galaxies in our sample is around
0.08.

We compute the ultraviolet spectral slopes using various definitions 
and compare the results to \betaGLX. We first consider the ultraviolet
spectral slope from the definition of \citet*{1994ApJ...429..582C},
which we denote by \betaCKS. This is obtained
from a power-law fit of the form $f_\lambda \propto \lambda^{\beta}$
(where $f_\lambda$ is the flux per unit wavelength) to the spectrum in
ten continuum bands that avoid strong stellar and interstellar 
absorption features (in particular, the 2175~{\AA} dust feature) in 
the range $1268\le \lambda \le2580$~\AA. We also consider the 
ultraviolet slopes derived from photometric observations with {\it 
ANS}, {\it OAO} and {\it UIT}. We compute the corresponding \betaANS,
\betaOAO\ and \betaUIT\ from continuum fits of the form $f_\lambda 
\propto \lambda^{\beta}$ to the broadband spectra obtained by 
convolving the {\it IUE} spectra with the filter response functions of
these satellites (Table~\ref{uvfilters}). 

\begin{table}
\caption{Effective wavelengths and FWHMs of the ultraviolet filters
used on-board different satellites.}
\label{uvfilters}
\centering
\begin{tabular}{lrr}
\hline
\hline
Satellite& $\lambda_e$ (\AA) &FWHM (\AA)\\
\hline
GALEX (FUV)& 1520 & 264 \\
GALEX (NUV)& 2310 & 795 \\
        &      &     \\
ANS (1) & 1550 & 149 \\
ANS (2) & 1950 & 149 \\
ANS (3) & 2200 & 200 \\
ANS (4) & 2500 & 150 \\
        &      &     \\
OAO (1) & 1550 & 240 \\
OAO (2) & 1910 & 260 \\
OAO (3) & 2460 & 380 \\
        &      &     \\
UIT (B1)& 1521 & 354 \\
UIT (A1)& 2488 &1147 \\ 
\hline
\end{tabular}
\end{table}

\begin{figure*}
\includegraphics[width=90mm,angle=-90]{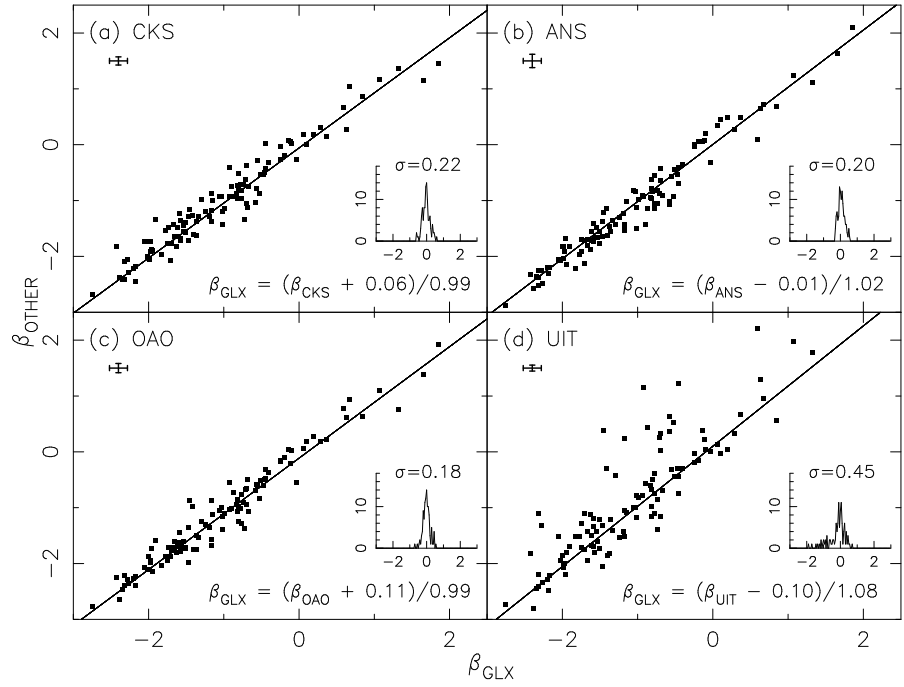}
\caption{Ultraviolet spectral slopes computed in the photometric
bands observed by different satellites plotted as a function of 
\betaGLX, for the 110 {\it IUE} galaxies with both far- and
near-ultraviolet spectra in the sample of Section\,2.1. In each
panel, the inset diagram shows the distribution of the difference
between the true value of \betaGLX\ and that obtained from the mean
linear relation (indicated at the bottom and drawn as a solid line)
for this sample, along with the 1$\sigma$ standard deviation.}
\label{figbeta}
\end{figure*}

Fig.~\ref{figbeta} shows \betaCKS, \betaANS, and \betaOAO\ and
\betaUIT\ as a function of \betaGLX\ for this sample. It can be
seen that \betaCKS, \betaANS\ and \betaOAO\ correlate extremely
well with \betaGLX. The correlation is worse for \betaUIT, mainly
because of the wider wavelength range of the near-ultraviolet 
filter of {\it UIT}, which extends redward to about 3060~{\AA}
(Table~\ref{uvfilters}). The mean linear relations between the 
ultraviolet slopes obtained from observations with different 
satellites and the reference ultraviolet slope \betaGLX\ are
drawn as solid lines in Fig.~\ref{figbeta} and are indicated at the 
bottom of each panel along with the 1$\sigma$ standard deviation. 
The tight correlations in Figs.~1(a)--1(c) allow us to estimate 
\betaGLX\ reliably from \betaCKS\, \betaANS\ and \betaOAO. Since
the relation is significantly worse for \betaUIT, we do not use
{\it UIT} data in the analysis described in the main text of the 
paper. 

The ultraviolet spectral slope \betaGLX\ can also be estimated
reliably for galaxies with only far-ultraviolet {\it IUE} spectra.
The reason for this is that the wavelength range of the far-ultraviolet
spectrograph on-board {\it IUE} encompasses 9 out of the 10 spectral 
windows originally used by \citet{1994ApJ...429..582C} 
to define \betaCKS.
In fact, for the 110 galaxies in the sample considered here, we find
that the ultraviolet slope \betaCKSshort\ obtained by fitting a power
law to the far-ultraviolet spectra in these 9 continuum bands 
correlates with \betaGLX\ as tightly as do \betaANS\ and \betaOAO. 
The mean linear relation is
\begin{equation}
\betaGLX\ = (\betaCKSshort - 0.38)/1.05\,,
\end{equation}
with a standard deviation $\sigma = 0.23$. 

We also adopt a common definition of the ultraviolet flux for 
galaxies observed with different satellites, $\Ffuv = \lambda_{\rm
FUV}\ffuv$. For those galaxies with {\it IUE} spectra, we compute 
the mean ultraviolet flux density $\overline {f}_{\rm FUV}$ using 
the response function of the {\it GALEX} far-ultraviolet filter 
($\lambda_{\rm FUV}=1520\,${\AA}). For those galaxies observed with
{\it OAO} and {\it ANS}, we compute $\overline {f}_{\rm FUV}$ by 
extrapolating the broadband spectrum from 1550~{\AA} (corresponding 
to the effective wavelength of the far-ultraviolet filters on-board 
both satellites) to 1520~{\AA}, using the ultraviolet slope (\betaOAO\
or \betaANS) measured for each galaxy. Given the narrow wavelength 
range over which the extrapolation is performed, this procedure does
not introduce any significant uncertainties in \ffuv\ and hence \Ffuv.

\label{lastpage}
\end{document}